\documentclass{iau} 
\usepackage{graphicx}
\usepackage{natbib}
\usepackage{amsmath}	
\title[Astroinformatics Challenges for Surveys] 
{Astroinformatics Challenges from Next-generation Radio Continuum Surveys}

\def\kms {\ifmmode{{\rm ~km~s}^{-1}}\else{~km~s$^{-1}$}\fi}
\def\lsun {\ifmmode{{\rm ~L}_\odot}\else{~L$_\odot$}\fi}
\def\deg {^{\circ} }
\def\sqdeg {\,deg$^2$}

\def\ujybm {\,$\mu$Jy/beam}

\newbox\grsign \setbox\grsign=\hbox{$>$} \newdimen\grdimen \grdimen=\ht\grsign
\newbox\simlessbox \newbox\simgreatbox
\setbox\simgreatbox=\hbox{\raise.5ex\hbox{$>$}\llap
 {\lower.5ex\hbox{$\sim$}}}\ht1=\grdimen\dp1=0pt
\setbox\simlessbox=\hbox{\raise.5ex\hbox{$<$}\llap
 {\lower.5ex\hbox{$\sim$}}}\ht2=\grdimen\dp2=0pt

\def\simless{\mathrel{\copy\simlessbox}}
\def \etal {\rm ~{\it \etal},~}
\def\apj {{\it Ap.~J.}}

\def\apjs {{\it Ap.~J.\ Suppl.}}
\def\aj {{\it A.~J.}}

\def\aap {{\it Astr.~Ap.}}

\def\mnras {{\it MNRAS}}

\def\pasa {{\it PASA}}

\author[Ray P. Norris]   
{Ray P. Norris$^{1,2}$,
}

\affiliation{
$^{1}$ Western Sydney University, Locked Bag 1797, Penrith South, NSW 1797, Australia\\email: {\tt raypnorris@gmail.com} \\[\affilskip]
$^{2}$ CSIRO Astronomy \& Space Science, PO Box 76, Epping, NSW 1710, Australia\\
}
\pubyear{2017}
\volume{325}  
\jname{Astroinformatics 2016}
\editors{Massimo Brescia, eds.}
\begin{document}

\maketitle

\begin{abstract}
The tens of millions of radio sources to be detected with next-generation surveys pose new challenges, quite apart from the obvious ones of processing speed and data volumes. For example, existing algorithms are inadequate for source extraction or cross-matching radio and optical/IR sources, and a new generation of algorithms are needed using machine learning and other techniques. The large numbers of sources enable new ways of testing astrophysical models, using a variety of ``large-n astronomy'' techniques such as statistical redshifts. Furthermore, while unexpected discoveries account for some of the most significant discoveries in astronomy, it will be difficult to discover the unexpected in large volumes of data, unless specific software is developed to mine the data for the unexpected.

\end{abstract}

\firstsection 
\section{Introduction}

The decade before the construction of the Square Kilometre Array \citep[SKA: ][]{dewdney09} has seen the construction of a number of SKA pathfinder telescopes, three of which  are already in operation: the  Meerkat telescope in South Africa \citep{jonas09}, the Australian SKA Pathfinder  \citep[ASKAP: ][]{johnston08}, and the Murchison Widefield Array in Australia \citep[MWA: ][]{tingay13}. Each of these three precursor telescopes is a major new instrument in its own right, using innovative new technology to exceed the performance of its predecessors by a significant margin.
\begin{figure}[h]
\begin{center}
\includegraphics[width=8cm, angle=0]{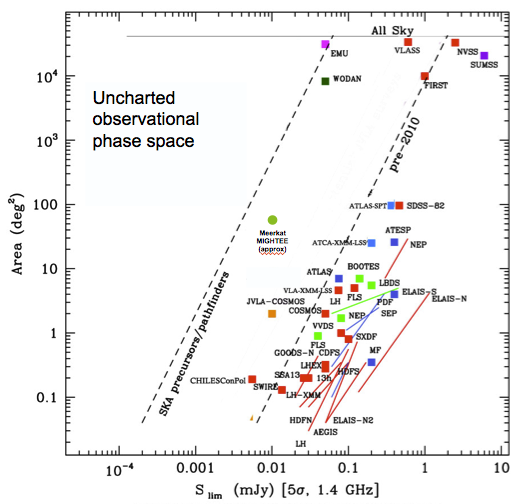}
\caption{Comparison of existing and planned deep 20 cm radio continuum surveys, based on a diagram by Isabella Prandoni. The horizontal axis shows the sensitivity, and the vertical axis shows the sky coverage. The right-hand diagonal dashed line shows the approximate envelope of existing surveys, which is largely determined by the availability of telescope time.  Surveys not at 20cm are represented at the equivalent 20 cm flux density, assuming a spectral index of -0.8.
The squares in the top-left represent the new radio surveys discussed in this paper.  
}
\label{surveys}
\end{center}
\end{figure}

For example,  Figure 1 shows the main 20cm radio surveys, both existing and planned. The largest existing radio survey, shown in the top right, is the wide but shallow NRAO VLA Sky Survey \citep[NVSS: ][]{condon98}. The most sensitive published radio survey is the deep but narrow JVLA-SWIRE (Lockman hole) observation in the lower left \citep{condon12}. 
Existing surveys are bounded by a diagonal line that roughly marks the limit of available  time on current-generation radio telescopes.  To the left of that line are three surveys (VLASS, JVLA-COSMOS, and CHILESConPol) that have been made possible by the  upgrade to the Jansky Very Large Array \citep[JVLA: ][]{moorsel14}. 

Even further to the left are three planned continuum surveys: WODAN, on the upgraded Westerbork telescope \citep{Rottgering10b}, the Meerkat surveys, whose exact specifications are still under discussion,
and  the Evolutionary Map of the Universe \citep[EMU: ][]{norris11}. Missing from this diagram are the important MWA, LOFAR \citep{haarlem13} and other surveys as they are at a very different frequency. 

All these surveys face similar challenges in terms of the volume of data, the large numbers of sources, and the resulting need to develop  tools to automate the data analysis that traditionally was done by hand. Recognising the need  to coordinate developments and avoid duplication of effort, the SKA Pathfinder Radio Continuum Survey Working Group (SPARCS) was established in 2010 and has had annual meetings since. Its achievements include the writing of a review paper listing these challenges \citep{Norris13} and the establishment of three reference survey fields at declination $\sim +30\deg, 0\deg$, and $-30\deg$, which can be observed by all existing and new radio telescopes, to ensure agreements on the measured positions, flux densities, polarisation, spectral index, etc. For example, the measured flux densities of sources in radio surveys are subject to a large number of subtle corrections and bias effects, that can cause measurement errors. 
The SPARCS fields have been chosen  to overlap with a field that is well-studied at other wavelengths, to maximise the  science to be obtained from these observations. 
It is planned to observe all three fields, as far and deeply as possible, with all existing survey telescopes as well as the new SPARCS surveys.  

The rest of this paper discusses the challenges common to all these surveys by discussing them in terms of the 
 largest of these surveys, EMU.
 
\subsection{ASKAP \& EMU}

The Australian SKA Pathfinder \citep[ASKAP: ][]{johnston08} is a new radio telescope nearing completion on the Australian SKA site in Western Australia, at the Murchison Radioastronomy Observatory. It consists of 36 12-metre antennas distributed over a region 6 km in diameter. It is revolutionary in that each antenna is equipped with a phased-array feed \citep[PAF: ][]{Bunton10} of 96 dual-polarisation pixels operating at 700--1800 MHz,
resulting in  a  field of view of 30 \sqdeg, and a much high survey speed than comparable single-pixel telescopes.

As well as producing images and source catalogues, the real-time processing pipeline \citep{Cornwell11} will also measure spectral index, spectral curvature, and all polarisation products.

ASKAP has a very high data rate: 70 Tbit/s from the antennas, and 220 TB/day from the correlator, which is  sent via an 800km dedicated fibre link to the Pawsey centre in Perth. There,  the processing software runs on a 200 Tflop/s Cray XC30 supercomputer that generates 70 PB/year of calibrated data, images, and catalogues.  The operating budget only allows 4 PB/year to be stored, so most of the spectral line time-series data is discarded.

ASKAP has already generated significant science 
 \citep[e.g. ][]{serra15, allison15, heywood16} in ``BETA'' and ``Early Science'' mode. Full operations are expected to start in early 2018, when the major survey programs, such as EMU, will commence.

ASKAP's all-sky continuum survey is EMU  \citep{norris11} which  will  survey 75\% of the sky to a sensitivity of 10\,\ujybm\ rms.
EMU will detect about 70 million galaxies, compared to the 2.5 million detected by all radio-telescopes over the entire history of radioastronomy. 
Not only will EMU have greater sensitivity than  previous large-area surveys, but it will also have high resolution and sensitivity to extended emission, and will
measure spectral index and, courtesy of the POSSUM project \citep{gaensler10}, polarisation for the strongest 
sources. 

EMU is driven by 18 ``Key Science Projects'' (KSPs), which cover areas ranging from cosmology and galaxy evolution through to Galactic science, but which are unified under the over-arching goal of understanding the evolution of the Universe. Importantly, one key science project is to discover the unexpected, which is discussed further below.
Many of the KSPs take advantage of the large number of sources available from EMU to operate in a ``large-n'' astronomy mode, where statistical inferences about populations of objects take precedence over the properties of individual objects.

EMU also has a number of ``Development Projects", covering challenges such as source extraction, classification,  cross-matching to multi-wavelength surveys, and redshift determination, in collaboration with other SPARCS members. 

Section 2 of this paper describes these challenges, Section 3 describes the methodology for discovering the unexpected, and Section 4 presents conclusions.

\section{Technical Challenges facing large radio continuum surveys}
\label{EVACAT}
\subsection{Compact Source Extraction and Measurement}

Astronomy has many tools available to find and extract sources in an image, and they are often used without checking their outputs.  However, none are suitable for an automated survey, as they require manual parameter adjustment, none accounts for variations in the point spread function across the image, and few sufficiently characterise the background and noise levels in an image  \citep[e.g.][]{huynh12}. There has been no systematic measurement of the reliability,
and false detection rate of the source finders, or the accuracy of their measurements of position, size, and flux density.

To ensure that the best source finders are used for EMU and other surveys, a number of well-known source finders (Aegean, Blobcat,  IMSAD,  PyBDSM, Selavy, Sextractor, and SFind) were pitted against each other in a data challenge by \cite{hopkins15}. The results were disappointing: even for isolated point sources, all current source finders fall well short of what should be theoretically achievable. In some cases sources well above the noise threshold were missed, and in other cases new sources were ``invented''. Measured source flux densities could be wrong by a factor of a few even for strong sources. The challenge of finding complex sources is even harder. 

The results of that data challenge have been used by the developers of the most successful radio-astronomical source finders (Aegean by \cite{ hancock11}, PyBDSM by \cite{mohan15} , and Selavy by \cite{whiting12} ) to enhance their software, and a new round of comparisons is now under way (K. Grieve et al., in preparation). The results of the new comparison will be used to ensure that EMU and other surveys use the best source finder available.

These analyses have also shown that  further development is essential to optimise and quantify the performance of current source-finding algorithms.

\subsection{Diffuse Source Extraction and Measurement}
\label{diffuse}
Whereas it was incorrectly thought that the compact source extraction problem had been solved, it is well known that the identification and extraction of diffuse emission is much harder. Astrophysically, diffuse emission is extremely important, for example as a tracer of cluster haloes, Galactic emission, and the cosmic web. 
However, previous small data volumes have been sufficiently small that work has focussed on hand-crafted tools \citep[e.g.][]{vernstrom15} and there has been little development of automated tools to detect diffuse emission. The thousands of cluster haloes expected to be detected by EMU means that the development of an automated tool is a high priority.
Several algorithms \citep{dabbech15, butler16, riggi16} are now under development for automatically detecting diffuse sources in radio-astronomical images.

\subsection{Classification and Cross-Identification}
\label{crossid}
Many science goals associated with radio surveys require the radio sources to be cross-identified with their counterparts, such as host galaxies, at other wavelengths. This is non-trivial since about 10\% of radio sources have several components.
For example,  two nearby unresolved radio components might either be the two lobes of a double radio source, or the radio emission from two  star-forming galaxies. Only by cross-identifying with optical/infrared data can these two cases be distinguished, since the star forming galaxies will have a host galaxy coincident with each of the radio components, whereas the host of the double radio source is likely to lie between them. Conversely, only by classifying the source do we know where to find the optical counterpart. Thus classification and cross-identification of radio sources are inextricably linked, and both processes must be performed simultaneously. 
Whilst this process is easy for the expert human, the 7 million complex sources expected to be detected by EMU pose a significant challenge, as there is currently no automated software to do this for large surveys.

Experience has shown that cross-identification with optical catalogues leads to a much higher false-identification rate than cross-identification with infrared (IR) catalogues, and so the primary cross-identification is  done with an IR catalogue, and then the IR position is used to cross-identify with other optical and IR catalogues. Several techniques are currently being evaluated, using the $\sim$5000 sources in the ATLAS data set \citep{norris06, franzen15} 
as a testbed, as follows:
\begin{itemize}
\item Manual classification and cross-identification to provide a training, test, and validation set (Jesse Swan et al., in preparation);
\item An extension of the likelihood-ratio approach \citep{sutherland92} is being developed (Stuart Weston et al., in preparation) to take into account the likelihood that several radio components may correspond to one optical component;
\item A Bayesian approach, which compares  the probabilities  of different classifications and cross-identifications, given a set of priors \citep{fan15};
\item The Radio Galaxy Zoo, in which thousands of citizen scientists do the classification and cross-identification by eye \citep{banfield16};
\item Several machine-learning projects, in which different machine learning algorithms such as neural nets are being evaluated to do the classification and cross-identification.
\end{itemize}
Finally, the results of all these techniques will be compared. It is likely that several of them will be used together, as described in \S\ref{wtf}.

\subsection{Redshifts}
\subsubsection{Statistical Redshifts}
For much of the science from radio continuum surveys, it is necessary to know the redshifts of the sources. However, only about 2\% of the 70 million EMU sources will have spectroscopic redshifts, and so a number of alternative approaches are being explored. For example, it is often sufficient to assign each source to a redshift bin, rather than measuring the  redshift of each source to a high precision. In many cases, knowledge of the redshift distribution of a subsample is sufficient, rather than redshifts of individual galaxies. I call such redshifts ``statistical redshifts''. For example, \cite{raccanelli14} show that EMU will be able to use the Integrated Sachs-Wolfe effect to measure primordial non-gaussianity to high accuracy. With no redshift information, the accuracy 
$\sigma(f_{NL})$ 
is uncompetitive, but  with only three redshift bins EMU can measure non-gaussianity more accurately than even Euclid.

\subsubsection{Radio Photometric Redshifts}

\begin{figure}
\begin{center}
\includegraphics[width=8cm, angle=0]{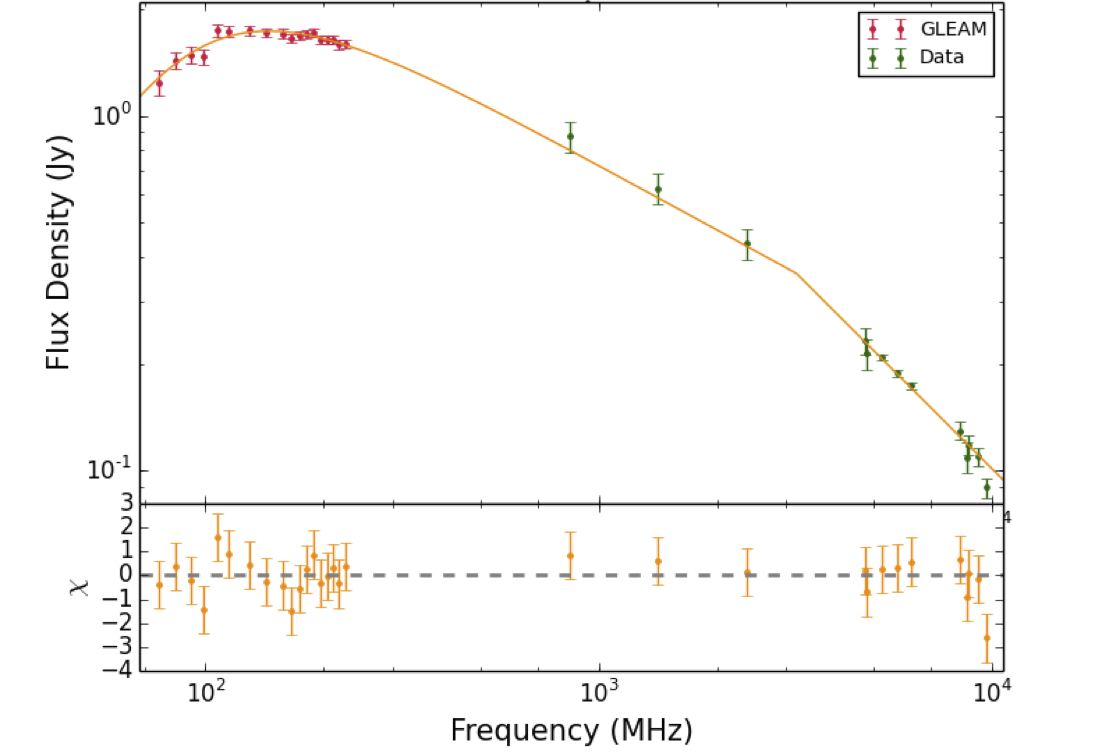}
\caption{The broad-band radio spectrum of a radio galaxy, from Jordan Collier et al (in preparation). Using either machine-learning photometric redshift techniques, or by modelling of the source, individual redshifts for some sources will be obtainable from radio data alone.
}
\label{collier}
\end{center}
\end{figure}

Radio sources were traditionally thought to have featureless power-law spectral-energy distributions (SED), and so could not be used to obtain redshifts directly. However, the advent of radio surveys at high and low frequencies have made it clear that a significant fraction of radio sources have rich radio SEDs such as that shown in Figure \ref{collier}, and so in principle may yield  photometric redshifts directly.

\subsubsection{Broadband Photometric Redshifts}

Conventional photometric redshifts are obtained by fitting SED templates to optical and infrared photometry measurements, using as many as 30 photometric bands \citep[e.g. ][]{salvato09}.
The requirements of radio continuum surveys are different from those driving these  techniques, in several respects:
\begin{itemize}
\item Precision is less important than the minimisation of the fraction of catastrophic outliers, with few applications requiring a precision $\Delta z < 0.1$. 
\item The available photometry,  mainly  from near-all-sky surveys such as WISE, VHS, and Skymapper, will be less homogenous that that used by (e.g.) \cite{salvato09}, with fewer photometry bands
\item EMU radio photometric radio data will be available for all sources, and spectral indices and  low-frequency radio data from MWA will be available for a significant fraction of sources. Any photometric code should make use of this radio  data.
\item Standard galaxy templates may not always be appropriate for radio sources, and tests \citep{salvato17} have shown that even simple machine learning techniques (e.g. kNN) can sometimes outperform template  techniques.
\item The use of machine learning techniques enables the use of other data, such as the polarisation data 
available for $\sim$10\% of EMU sources. Detection of polarisation virtually guarantees the source is an AGN.
\end{itemize}

\subsubsection{Spatial Clustering Redshifts}
\cite{rahman16} and others have shown that redshifts can be estimated from spatial clustering information of nearby galaxies. This is harder for radio continuum surveys, because radio surveys generally extend to much higher redshifts than optical surveys, making it difficult to obtain a training set co-located with the target set.  For example, the median redshift for EMU sources is  $z \sim 1.4$. So while a training set for photometric redshift techniques may be obtained from one well-studied part of the sky, a training set for spatial clustering will be available only for low ( $z \simless 1$) redshift sources.

\section{WTF? Discovering the Unexpected}
\label{wtf}
\subsection{The Process of Discovery}

\begin{figure}
\includegraphics[width=8cm, angle=0]{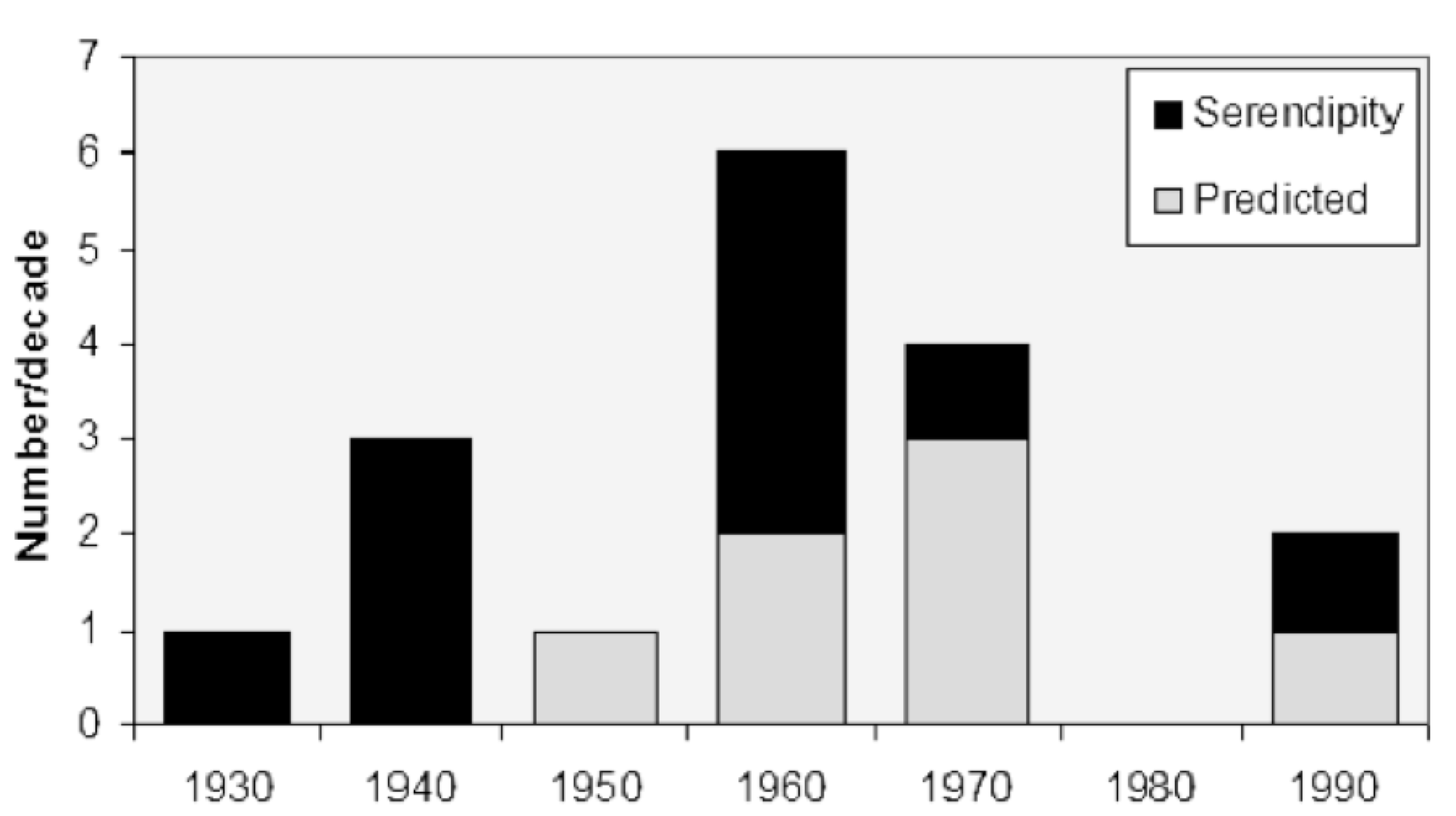}
\caption{A plot of recent major astronomical discoveries, taken from Ekers (2009)
of which seven were ``known-unknowns'' (i.e. discoveries made by testing a  prediction) and ten were ``unknown-unknowns'' (i.e.. a serendipitous result found by chance while performing an experiment with different goals). }
\label{ekers}
\end{figure}

At least half the major discoveries in astronomy are unexpected \citep{harwit81, wilkinson04, wilkinson07, kellermann09, ekers09, wilkinson15}.
For example, \cite{ekers09} examined  17 major astronomical discoveries in the last 60 years, and  concluded (see Figure \ref{ekers}) that only seven resulted from systematic observations designed to test a hypothesis or probe the nature of a type of object. The remaining ten unexpected discoveries resulted either from new technology, or from observing the sky in an innovative way, exploring uncharted parameter space.   Similarly, \cite{norris17} showed that, of the 10 greatest discoveries made with the Hubble Space telescope, only one (using Cepheids to measure the Hubble constant) was amongst its science goals. The other nine, including the discovery of dark energy, were in some sense unexpected.

A prime example of an unexpected discovery was the discovery of pulsars by Jocelyn Bell. She  observed the radio sky for the first time with high time resolution, to study interstellar scintillation, and thereby explored an unexplored part of observational parameter space. She found ``bits of scruff'' on the chart recorder, which she realised could not be due to terrestrial interference, but represented a new type of astronomical object. She describes the process in detail in \cite{bell09}.

Figure 1 shows that radio continuum surveys such as EMU are also venturing into unexplored parts of observational parameter space. From Occam's razor, the unexplored region of observational parameter space to the left of the line presumably contains as many  potential new discoveries per unit parameter-space as the region to the right. 
EMU should therefore significantly expand the volume of observational parameter space, so  in principle should discover unexpected new phenomena and new types of object. 

However, it's unlikely that a latter-day Jocelyn Bell could discover the unexpected in ASKAP data.
She discovered pulsars by laboriously sifting through all her data, and noticing a tiny anomaly that didnÕt fit her understanding of the telescope. If she were observing with ASKAP, she would have to sift through petabytes of data from a machine that is so complex that nobody truly understands every bit of it. 

The only way of extracting science from large volumes of data is to interrogate the data with a well-posed question, such as `plot the specific cosmic star formation rate of star-forming galaxies as a function of redshift'.
 This is a very efficient way of answering the Òknown-unknownsÓ, but it is incapable of finding the Òunknown-unknownsÓ. 
 Since the human brain cannot sift through petabytes by eye, then we must rely on tools to detect the unexpected, and such tools do not currently exist.

We have therefore started a  project within EMU, named Widefield ouTlier Finder, or WTF, to develop techniques for mining large volumes of astronomical data for the unexpected, using  machine-learning techniques and algorithms. There are two types of unexpected discovery: the discovery of unexpected objects, and the discovery of unexpected phenomena, which may appear as an anomaly in the properties of a sample of objects. A detailed description is given by \cite{norris17}.

\subsection{WTF1: Discovering unexpected objects}
\begin{figure}[h]
\includegraphics[width=12cm, angle=0]{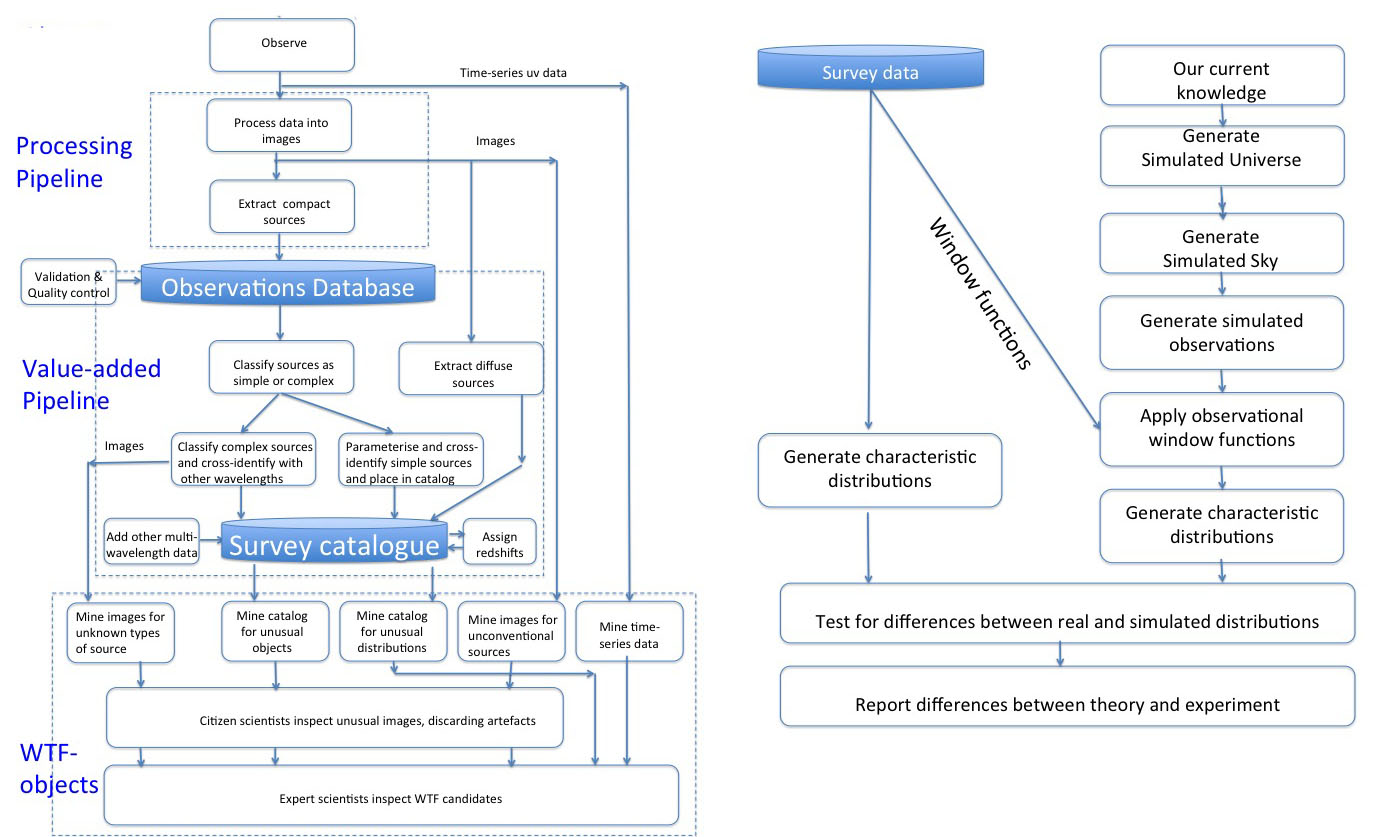}
\caption{(Left) The flowchart for discovering unexpected objects. (Right)The flowchart for discovering unexpected phenomena.
}
\label{flowchart}
\end{figure}

There are currently  $\sim$ 2.5 million known radio sources. 
EMU is expected to detect about 70 million objects, and so there is a good chance that these will include new  classes of radio source. WTF will search for them using the process shown
 in Figure \ref{flowchart}. Although this is designed for EMU,  the broad approach is applicable to any survey.

The
ASKAPSOFT real-time processing pipeline processes EMU data to produce images and source catalogues in the ``CASDA'' Observation database. This is followed by the ``Value-added pipeline'',  which implements the processes described in \S\ref{EVACAT} to produce the Survey catalogue, EVACAT (EMU Value-added Catalogue). WTF then mines EVACAT, together with the radio images, for unexpected objects.

One function of WTF is to mine the images for unconventional sources. For example,  a ring of emission several arcmin in diameter but with an amplitude of only half the rms noise level in any one pixel, would be invisible to the human eye, or to a conventional source extraction code. Such a ring could easily be detected using a  suitable matched filter, such as a Hough transform \citep{hollitt12}. Many other examples of potential diffuse and unconventional sources may be imagined.
Detecting sources with unknown unconventional morphology is much harder and is the subject of continuing research such as  \cite{geach12} and \cite{baron16}.

The catalogue will be searched in an n-dimensional parameter space with axes such as flux density, spectral index, and IR-to-radio ratio.  Known types of object (e.g. stars, galaxies, quasars) will appear as clusters in this parameter space. Algorithms are being explored that will search the parameter space for anomalies, such as clusters of objects that do not correspond to known types of objects.  

\subsection{WTF2: Discovering Unexpected Phenomena}

An unexpected discovery can result from the properties of a sample of objects differing from those predicted by theory. For example, dark energy was discovered \citep{riess98, perlmutter99} when the relationship between the brightness and redshift of type 1A supernovae  differed from that predicted by theory. Here I describe an approach in which  data is tested against theory to reveal discrepancies. Rather than trying to derive theoretical quantities from the data, which requires a number of assumptions and corrections, the opposite approach is taken of generating simulated observations from the theory, which can then easily have the observational constraints (the ``window function'') applied, and then compared to data in an empirical ``characteristic distribution'' (e.g. source counts as a function of observed flux density, or the angular power spectrum).

By doing this, the simulations are effectively being used to encapsulate our current understanding of astrophysics, so that we can check if the observed data is consistent with our current understanding. Any significant difference between the two either represents an error in the data or simulation, or an unexpected discovery.
%
%
This process is shown in Figure \ref{flowchart}, and includes the following steps. 
The starting point is a simulation, such as the Millennium Simulation \citep{springel05} which encapsulates our knowledge about cosmology and galaxy formation. From this is generated a simulated sky, using our knowledge of the observed properties of galaxies, using a tool such as the Theoretical Astrophysical Observatory \citep[TAO: ][]{bernyk16}, together with a  semi-empirical model of radio sources. 
The model sky is them converted to a simulated observed sky using observational constraints such as sensitivity and resolution. The sky is then ``observed'' using the window function of the real data, which includes factors such as area of sky observed, and any varying sensitivity across the observations.

\begin{figure}
\begin{center}
\includegraphics[width=8cm, angle=0]{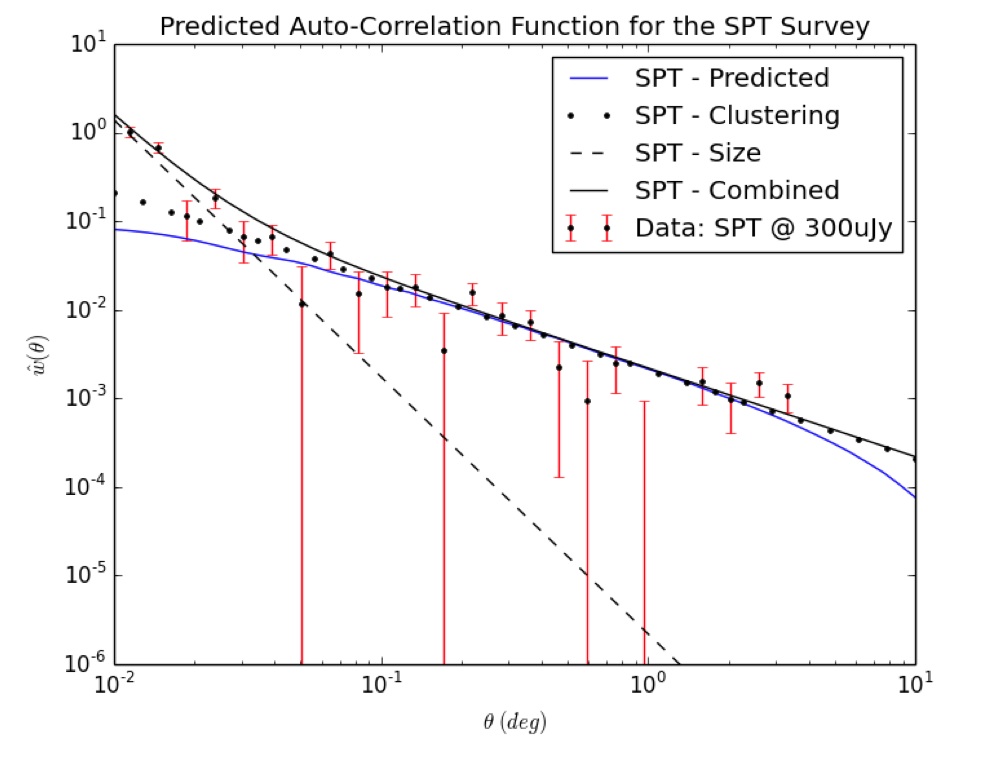}
\caption{The angular power spectrum for radio sources in the SPT field, taken from Rees et al (2017). Points with error bars are the measured angular power spectrum of the data obtained by O'Brien et al (2017), and the blue line shows the distribution predicted by the semi-empirical model described in the text. The dotted line shows the cosmological signal predicted by LCDM,and the dashed line show the effect of radio source size and double radio sources. The solid black line is the sum of these latter two predictions.}
\label{rees}
\end{center}
\end{figure}

For example, Figure \ref{rees}  taken from \cite{rees17}, shows the angular power spectrum for radio sources in the SPT (South Pole Telescope)  field, using the radio observations described by \cite{obrien}. The simulated data were based on the Millennium Simulation, from which a simulated sky of galaxies was generated using the TAO  tool. From this, a radio sky was generated using semi-empirical assumptions about the properties of radio sources based on the zFOURGE survey \citep{rees16}. 

\section {Conclusions}

Next-generation radio continuum surveys face a number of technical challenges and opportunities, which must be addressed if we are to extract the science from the tsunami of data. Many of the solutions will take advantage of the large data volumes and use machine learning algorithms. Over half the major discoveries in astronomy are unexpected, but  are unlikely to be made by humans in the large data volumes that characterise next-generation surveys. Instead, software must be designed explicitly  to maximise their ability to mine the data for unexpected discoveries, including both unexpected objects and unexpected phenomena.

\end{document}